\documentclass[a4paper,12pt]{article}

\usepackage{authblk}
\usepackage{hyperref}
\hypersetup{colorlinks=true, linkcolor=blue, citecolor=blue, urlcolor=blue}
\usepackage{braket}
\usepackage{amssymb}
\usepackage{amsmath}
\usepackage{amsthm}
\usepackage{amscd}
\usepackage{graphicx}
\usepackage{placeins}
\usepackage{geometry}
\usepackage{bm, physics}
\usepackage{mathtools}

\theoremstyle{definition}

\theoremstyle{remark}

\begin{document}

\title{Semiclassical Ehrenfest Paths in Open Quantum Systems}

\author[1]{Xiao-Kan Guo\thanks{E-mail: kankuohsiao@whu.edu.cn}}
\affil[1]{School of Mathematics and Physics, Yancheng Institute of Technology, Jiangsu 224051, China}

\date{\today}

\maketitle

\begin{abstract}
We study the semiclassical Ehrenfest trajectories in open quantum systems.
Using the Fokker--Planck equation for the Gaussian mixing measure, we embed the generalized Ehrenfest theorem recently obtained for open quantum systems into this phase-space picture and study the time evolution of the expectations of observables with respect to the Gaussian mixture.
We show how the coherent and irreversible contributions are separated in the frictionless harmonic regime relevant to the semiclassical approximation.
Our work provides a transparent phase-space interpretation of the emergence of classical trajectories in open quantum dynamics.
\end{abstract}

\section{Introduction}
\label{sec:intro}

The Ehrenfest theorem \cite{Ehrenfest1927} provides a well-known connection between quantum and classical dynamics by relating the time derivatives of expectation values of position and momentum to the expectation value of the force.
The formal similarity between the time evolution of quantum expectation values and the time evolution in Newton's second law, however, cannot be considered as an identity between the quantum and classical dynamics, as the expectation of the force is not generically equal to the force evaluated at the expectation of the position. To find a more direct connection between the quantum and classical dynamics, one could resort to the semiclassical approximation of quantum dynamics using, for example, the thawed Gaussian \cite{Heller1975} or the general semiclassical evolution of wave packets \cite{L86}, as the semiclassical wave packet. Considering the time evolution of the expectations of observables in such semiclassical wave packets, one is led to the semiclassical trajectories for the wave packets \cite{Liberalquino2013}, which provides a more direct link between the quantum and classical dynamics than the original Ehrenfest theorem.

For open quantum systems, however, the above-mentioned connections break down because of the environmental effects such as dissipation and decoherence. To take these effects into account, we should seek for the modified versions of the Ehrenfest theorem and the semiclassical Ehrenfest trajectories adapted to the open quantum systems.
A generalization of the Ehrenfest theorem to open quantum systems has been recently proposed \cite{Vallejo2025}, revealing additional contributions from entropy changes and coherence. Ehrenfest-type theorems for open systems with unbounded generators and observables have also been studied in the quantum Bernoulli noise framework \cite{Chen2025}, where the emphasis is on the well-posedness of the underlying master equation.
On the other hand, the semiclassical approximation method based on Gaussian states has also been applied to open quantum systems \cite{Graefe2018}.

Notice that, in \cite{Graefe2018}, the authors studied the time evolution of Gaussian states and their superpositions under the Lindblad dynamics. Since for an open quantum system the internal energy, as the quantum expectation value of the Hamiltonian operator, is no longer conserved, it should be accompanied by the thermodynamic exchange of heat and work with the environment. In order to reproduce such a thermodynamic behavior at the semiclassical level, we need to consider an ensemble of Gaussian states instead of a single wave packet. To this end, we can consider the \emph{Gaussian mixture} representation, which is a rigorous time-dependent semiclassical mixture framework introduced in \cite{Hernandez2025} for general Lindblad evolutions. Recently,  \cite{Galkowski2025} established a rigorous long-time classical--quantum correspondence for Lindblad evolution that for at most quadratically growing Hamiltonians and at most linearly growing jump functions satisfying ellipticity assumptions, the quantum observable remains close to the classical Fokker--Planck evolution in Hilbert--Schmidt norm for times vastly exceeding the Ehrenfest time, and their Theorem~2 covers mixtures of coherent states with fixed covariance. In contrast, the present work allows general Gaussian states whose covariance evolves under the local harmonic flow, uses the Fokker--Planck equation for the mixing measure (following \cite{Hernandez2025}), and connects the Gaussian-mixture evolution to the generalized Ehrenfest theorem \cite{Vallejo2025}. We show how the symplectic flow of the Gaussian components and the diffusion of the mixing measure correspond, respectively, to the coherent and irreversible contributions in that theorem, in the frictionless harmonic regime in which the local harmonic approximation is exact. In this way, the present work extends the earlier semiclassical Ehrenfest paths \cite{Liberalquino2013} to open systems.

We begin in Sec.~\ref{sec:framework} by introducing the basic definitions including the Gaussian mixture representation and the definition of a single Gaussian state's semiclassical path under the local harmonic approximation.
Then, in Sec.~\ref{subsec:mu-evolution}, we give a detailed derivation of the Fokker--Planck equation for the mixing measure.
Finally, in Sec.~\ref{subsec:generalized-ehrenfest} we demonstrate how our construction naturally recovers the generalized Ehrenfest theorem \cite{Vallejo2025}. We illustrate the formalism with an example in Sec.~\ref{sec:example}, and Sec.~\ref{sec:conclusion} concludes.

\section{Theoretical Framework}
\label{sec:framework}
Our starting point is the representation of a large class of quantum states $\rho(t)$ as a statistical mixture of Gaussian wave packets. This idea has deep roots in semiclassical analysis \cite{Heller1975} and has recently been placed on a rigorous footing for open quantum systems \cite{Hernandez2025}.

Consider a $2d$-dimensional phase space $\mathbb{R}^{2d}$ with coordinates $(x^1,\dots,x^d,p_1,\dots,p_d)$.
A {Gaussian quantum state} \(\hat{\tau}_{\alpha,\sigma}\) is a density operator on the Hilbert space \(L^2(\mathbb{R}^d)\) whose Wigner function \(\mathcal{W}_\hbar[\hat{\tau}_{\alpha,\sigma}]\) is a Gaussian distribution on the phase space \(\mathbb{R}^{2d}\).
It is characterized by its mean, or centroid \(\alpha\in\mathbb{R}^{2d}\), and its covariance matrix \(\sigma\), a real symmetric positive definite \(2d\times 2d\) matrix satisfying the uncertainty principle \(\sigma + \frac{i\hbar}{2}\omega \ge 0\), where \(\omega\) is the standard symplectic form on $\mathbb{R}^{2d}$.
The Wigner function $\mathcal{W}_\hbar[\hat{\tau}_{\alpha,\sigma}](\alpha+\beta)$ reads
\begin{align}
\mathcal{W}_\hbar[\hat{\tau}_{\alpha,\sigma}](\alpha+\beta)& \equiv \tau_{\alpha,\sigma}(\alpha+\beta) = \nonumber\\
&=\frac{1}{(2\pi)^d \sqrt{\det\sigma}} \exp\!\left(-\frac{1}{2}\beta^{\mathsf{T}}\sigma^{-1}\beta\right),\qquad \beta\in\mathbb{R}^{2d}.
\end{align}
When the covariance \(\sigma\) satisfies the additional symplectic condition \(\sigma\omega\sigma = \frac{\hbar^2}{4}\omega\), the state is pure; otherwise it is mixed and can be decomposed as a convex combination of pure Gaussian states.

We approximate the exact density matrix \(\rho(t)\) by a Gaussian mixture \(\tilde{\rho}(t)\) defined as
\begin{equation}
\label{eq:mixture}
\tilde{\rho}(t) := \iint_{\mathbb{R}^{2d}\times\mathcal{S}} \hat{\tau}_{\alpha,\sigma}\; \mu_t(\mathrm{d}\alpha,\mathrm{d}\sigma),
\end{equation}
where \(\mathcal{S}\) denotes the set of admissible covariance matrices (ensuring \(\hat{\tau}_{\alpha,\sigma}\) is a valid quantum state), and \(\mu_t\) is a time-dependent probability measure on \(\mathbb{R}^{2d}\times\mathcal{S}\) satisfying \(\iint\mu_t = 1\). For the purposes of the semiclassical approximation, we restrict the measure to the pure not-too-squeezed set $S_{\mathrm{NTS}}(\xi)$ defined below, as in \cite{Hernandez2025}. For an initial coherent state \(\hat{\tau}_{\alpha_0,\sigma_*}\) with \(\sigma_* = \frac{\hbar}{2}\mathbb{I}_{2d}\), we can choose \(\mu_0(\mathrm{d}\alpha,\mathrm{d}\sigma)=\delta(\alpha-\alpha_0)\delta(\sigma-\sigma_*)\) so that \(\tilde{\rho}(0)=\rho(0)\) exactly. In the following we will refer to $\mathbb{R}^{2d}\times\mathcal{S}$ as the extended phase space with coordinates $(\alpha,\sigma)$.

A probability mixture of Gaussian states has a nonnegative Wigner function. Consequently, the representation is exact for Gaussian states and their probability mixtures, and it cannot approximate states with negative Wigner functions, such as odd Fock states \cite{Hudson1974}. A general state would require a signed measure. In this work we restrict ourselves to probability measures, as in \cite{Hernandez2025}. Within this class, the Gaussian mixture representation is both flexible and convenient. It is natural in the Wigner-measure framework \cite{Lions1993}, and the Gaussian components remain Gaussian under harmonic approximations to the dynamics, as we now show.

Following the strategy of \cite{Hernandez2025}, consider an open quantum system governed by the Lindblad master equation
\begin{equation}
\partial_t \hat{\rho} = \hat{\mathcal{L}}[\hat{\rho}] = -\frac{i}{\hbar}[\hat{H},\hat{\rho}] + \frac{1}{\hbar}\sum_k \Big( \hat{L}_k\hat{\rho}\hat{L}_k^\dagger - \frac{1}{2}\{\hat{L}_k^\dagger\hat{L}_k,\hat{\rho}\} \Big),
\end{equation}
where the Hamiltonian operator $\hat{H}$ and the Lindbladian generator $\hat{L}_k$ have their respective Weyl symbols \(H(\alpha)\) and \(L_k(\alpha)\) on the phase space $\mathbb{R}^{2d}$.
Define \(\hat{M}_k := \hat{L}_k - L_k(\alpha)\).
We expand all the symbols to second order around the centroid \(\alpha\), and the resulting {local harmonic Lindbladian} generator \(\hat{\mathcal{L}}^{(\alpha)}\) is given by
\begin{align}
\hat{\mathcal{L}}^{(\alpha)}[\hat{\rho}] = &-\frac{i}{\hbar}\Big[ \hat{H}^{[\alpha,2]} + \operatorname{Im}\sum_k L_k(\alpha)\hat{M}_k^{[\alpha,2]\dagger},\hat{\rho}\Big] +\nonumber\\
&+ \frac{1}{\hbar}\sum_k\Big( \hat{M}_k^{[\alpha,1]} \hat{\rho} \hat{M}_k^{[\alpha,1]\dagger}
   - \frac{1}{2}\{\hat{M}_k^{[\alpha,1]\dagger}\hat{M}_k^{[\alpha,1]},\hat{\rho}\} \Big),
\end{align}
where \(\hat{H}^{[\alpha,2]}\) and \(\hat{M}_k^{[\alpha,2]}\) are the Weyl quantizations of the truncated Taylor series of the Weyl symbols to the second order in $\alpha$. This generator is harmonic (with quadratic Hamiltonian and linear Lindbladian operators) and therefore maps Gaussian states to Gaussian states.

Applying \(\hat{\mathcal{L}}^{(\alpha)}\) to a Gaussian state whose centroid coincides with the expansion point \(\alpha\) yields the following equations for the centroid \(\alpha(t)\) and the covariance matrix \(\sigma(t)\) \cite{Hernandez2025}:
\begin{align}
 \frac{d\alpha^a}{dt} &= U^a(\alpha)  \label{eq:alpha-evol}\\
 \frac{d\sigma^{ab}}{dt} &= S^{ab}(\alpha,\sigma)  \label{eq:sigma-evol}
\end{align}
where
\begin{align}
U^a(\alpha) &:= \partial^a H(\alpha) + G^a(\alpha), \qquad
G^a(\alpha) := \operatorname{Im}\sum_k L_k(\alpha)\,\partial^a L_k^*(\alpha), \label{eq:drift}\\
S^{ab}(\alpha,\sigma) &:= [F^a_{\,c}(\alpha)+\Gamma^a_{\,c}(\alpha)]\sigma^{cb} + \sigma^{ac}[F^b_{\,c}(\alpha)+\Gamma^b_{\,c}(\alpha)] + D^{ab}(\alpha), \label{eq:S}\\
F^a_{\,b}(\alpha) &:= \partial_b\partial^a H(\alpha), \qquad
\Gamma^a_{\,b}(\alpha) := \partial_b G^a(\alpha), \qquad
D^{ab}(\alpha) := \hbar\,\operatorname{Re}\sum_k (\partial^a L_k)(\partial^b L_k^*)(\alpha). \nonumber
\end{align}
The indices \(a,b,\ldots\) run over the \(2d\) phase-space coordinates. We raise and lower indices with the symplectic form $\omega$: $\partial^a := \omega^{ab}\partial_b$, $\partial_a := \partial/\partial\alpha^a$, with the sign convention $\omega^{xp}=+1$, so that $\partial^x=\partial_p$ and $\partial^p=-\partial_x$. This is the same convention as in \cite{Hernandez2025}.
Equation \eqref{eq:alpha-evol} shows that the centroid follows a classical trajectory modified by the friction vector \(G^a\). Equation \eqref{eq:sigma-evol} describes how the Gaussian packet is stretched by the Hamiltonian flow (through \(F\)) and friction (through \(\Gamma\)), and broadened by diffusion (through \(D\)). The diffusion matrix \(D\) is proportional to \(\hbar\), reflecting its quantum origin. Together, \((\alpha(t),\sigma(t))\) constitute the {semiclassical path} of the Gaussian component.

For any observable $\hat{O}$, its expectation value in the Gaussian state \(\hat{\tau}_{\alpha,\sigma}\) (with fixed $\alpha$ and $\sigma$)  is given by
\begin{equation}\label{999}
\langle \hat{O} \rangle_{\tau_{\alpha,\sigma}} = \int \tau_{\alpha,\sigma}(\beta)\, O(\beta)\, d\beta,
\end{equation}
where $O(\beta)$ is the Weyl symbol of $\hat{O}$. This integral \eqref{999} is a Gaussian average weighted by \(O(\beta)\).
In particular, the first $d$ components of the centroid are the expectation values of $\hat{x}^a$, and the last $d$ components are those of $\hat{p}_a$.

\section{Evolution Equation for the Mixing Measure \(\mu_t\)}
\label{subsec:mu-evolution}
We now study how the mixing measure \(\mu_t\) must evolve so that the mixture \(\tilde{\rho}(t)\) faithfully approximates the exact Lindblad dynamics.
 The following derivation makes explicit the transport equation used in \cite{Hernandez2025}.

Because the parameters $(\alpha,\sigma)$ of each individual Gaussian evolve according to the deterministic system of ordinary differential equations \eqref{eq:alpha-evol} and \eqref{eq:sigma-evol}, the natural starting point is the \emph{Lagrangian} description via the push-forward of the initial measure.
Let \(\Phi_t:(\alpha_0,\sigma_0)\mapsto (\alpha(t),\sigma(t))\) be the flow generated by \eqref{eq:alpha-evol} and \eqref{eq:sigma-evol}. Given an initial probability measure \(\mu_0\) on \(\mathbb{R}^{2d}\times\mathcal{S}\), we define the time-dependent measure \(\mu_t\) as the push-forward under this flow:
\begin{equation}
\label{eq:pushforward}
\mu_t(B) := \mu_0\big(\Phi_t^{-1}(B)\big)
\end{equation}
for any measurable set \(B\subset\mathbb{R}^{2d}\times\mathcal{S}\). Equivalently, for any smooth compactly supported test function \(f(\alpha,\sigma)\), we have
\begin{equation}\label{10}
\iint f(\alpha,\sigma)\,\mu_t(\mathrm{d}\alpha,\mathrm{d}\sigma)
= \iint f\big(\Phi_t(\alpha_0,\sigma_0)\big)\,\mu_0(\mathrm{d}\alpha_0,\mathrm{d}\sigma_0).
\end{equation}
The mixture \(\tilde{\rho}(t)\) defined in \eqref{eq:mixture} then becomes
\begin{equation}
\tilde{\rho}(t)
= \iint \hat{\tau}_{\alpha,\sigma}\,\mu_t(\mathrm{d}\alpha,\mathrm{d}\sigma)= \iint \hat{\tau}_{\Phi_t(\alpha_0,\sigma_0)}\,\mu_0(\mathrm{d}\alpha_0,\mathrm{d}\sigma_0).
\end{equation}

Differentiating the relation \eqref{10} with respect to time gives,
\begin{align}
\frac{d}{dt} \iint f(\alpha,\sigma)\,\mu_t(\mathrm{d}\alpha,\mathrm{d}\sigma)
&= \frac{d}{dt} \iint f\big(\Phi_t(\alpha_0,\sigma_0)\big)\,\mu_0(\mathrm{d}\alpha_0,\mathrm{d}\sigma_0)= \nonumber\\
&= \iint \Big[ \frac{d\alpha^a}{dt}\frac{\partial f}{\partial\alpha^a} + \frac{d\sigma^{ab}}{dt}\frac{\partial f}{\partial\sigma^{ab}} \Big]_{\Phi_t(\alpha_0,\sigma_0)} \mu_0(\mathrm{d}\alpha_0,\mathrm{d}\sigma_0) =\nonumber\\
&= \iint \Big[ U^a(\alpha) \frac{\partial f}{\partial\alpha^a} + S^{ab}(\alpha,\sigma) \frac{\partial f}{\partial\sigma^{ab}} \Big] \mu_t(\mathrm{d}\alpha,\mathrm{d}\sigma). \label{eq:weak-transport}
\end{align}
If the measure \(\mu_t\) possesses a smooth density (also denoted by \(\mu_t(\alpha,\sigma)\)) with respect to the Lebesgue measure \(\mathrm{d}\alpha\,\mathrm{d}\sigma\), then integration by parts in \eqref{eq:weak-transport} yields
\begin{equation}
\partial_t \mu_t = -\frac{\partial}{\partial\alpha^a}\big(U^a \mu_t\big) - \frac{\partial}{\partial\sigma^{ab}}\big(S^{ab} \mu_t\big). \label{eq:naive-transport}
\end{equation}

Notice that the diffusion $D^{ab}(\alpha)$ in $S^{ab}$ requires a more careful treatment of the equation \eqref{eq:naive-transport}.
A crucial observation is that the derivative of $\tau_{\alpha,\sigma}$ with respect to the covariance matrix can be expressed as a second derivative with respect to the centroid (see Appendix~B.1 of \cite{Hernandez2025}),
\begin{equation}
\label{eq:sigma-alpha-relation}
\frac{\partial \tau_{\alpha,\sigma}}{\partial \sigma^{ab}} = \frac{1}{2} \frac{\partial^2 \tau_{\alpha,\sigma}}{\partial \alpha^a \partial \alpha^b}.
\end{equation}
Because trace and integration against a symbol commute with these derivatives, the same relation holds for the density operator:
\begin{equation}\label{15}
\frac{\partial \hat{\tau}_{\alpha,\sigma}}{\partial \sigma^{ab}} = \frac{1}{2} \frac{\partial^2 \hat{\tau}_{\alpha,\sigma}}{\partial \alpha^a \partial \alpha^b}.
\end{equation}
Now, by letting $f(\alpha,\sigma)= \hat{\tau}_{\alpha,\sigma}$, we obtain from \eqref{eq:weak-transport} that
\begin{equation}\label{16}
\partial_t \tilde{\rho}(t) = \iint \Big( U^a \partial_{\alpha^a}\hat{\tau} + S^{ab} \partial_{\sigma^{ab}}\hat{\tau}_{\alpha,\sigma}\Big) \mu_t(\mathrm{d}\alpha\mathrm{d}\sigma).
\end{equation}
We now decompose the matrix \(S^{ab}(\alpha,\sigma)\) into a part that preserves the pure-state character (i.e., symplectic evolution) and a part that can be interpreted as diffusion:
\begin{equation}
\label{eq:S-decomp}
S^{ab}(\alpha,\sigma) = S_0^{ab}(\alpha,\sigma) + S_{\mathrm{D}}^{ab}(\alpha,\sigma),
\end{equation}
where \(S_{\mathrm{D}}\) is chosen to be symmetric and positive semidefinite. The precise algebraic construction of such a decomposition under the ``not-too-squeezed'' condition\footnote{The local harmonic approximation is accurate only as long as the Gaussian states remain sufficiently localized, i.e., not too squeezed in any phase-space direction.
Given a squeezing ratio \(\xi \in (0,1]\), a pure Gaussian state \(\hat{\tau}_{\alpha,\sigma}\) is called \emph{not too squeezed} if its covariance matrix satisfies
$
\sigma \ge \xi \frac{\hbar}{2} \mathbb{I}_{2d}.
$
The decomposition $S=S_0+S_D$ used below is the one constructed in Lemma~7.1 of \cite{Hernandez2025}: $S_0$ is the purity-preserving (symplectic) part and $S_D\ge0$ is the diffusion part. In the frictionless case one may take $S_D=D$ and $S_0=S-D$.
}
 is given in Lemma~7.1 of \cite{Hernandez2025}. For our purposes it suffices that such a decomposition exists and satisfies the properties stated in that lemma. With this decomposition, the derivative along the flow acting on a Gaussian state becomes
\begin{equation}
\frac{d}{dt} \hat{\tau}_{\alpha,\sigma}
= U^a \frac{\partial \hat{\tau}}{\partial\alpha^a} + (S_0^{ab}+S_{\mathrm{D}}^{ab}) \frac{\partial \hat{\tau}}{\partial\sigma^{ab}} = U^a \frac{\partial \hat{\tau}}{\partial\alpha^a} + S_0^{ab} \frac{\partial \hat{\tau}}{\partial\sigma^{ab}} + \frac{1}{2} S_{\mathrm{D}}^{ab} \frac{\partial^2 \hat{\tau}}{\partial\alpha^a\partial\alpha^b}, \label{eq:dt-tau-decomposed}
\end{equation}
where the last term is due to the identity \eqref{15}. With this decomposition, \eqref{16} becomes
\begin{equation}
\partial_t \tilde{\rho}(t) = \iint \Big( U^a \partial_{\alpha^a}\hat{\tau} + S_0^{ab} \partial_{\sigma^{ab}}\hat{\tau} + \frac{1}{2} S_{\mathrm{D}}^{ab} \partial_{\alpha^a}\partial_{\alpha^b}\hat{\tau} \Big) \mu_t(\mathrm{d}\alpha\mathrm{d}\sigma).
\end{equation}
In the construction of the mixture, we can absorb the second-derivative term into the evolution of the mixing measure \(\mu_t\) by performing an integration by parts:
\begin{equation}
\partial_t \tilde{\rho}(t) = \iint \hat{\tau}_{\alpha,\sigma} \Big[ -\frac{\partial}{\partial\alpha^a}\big(U^a \mu_t\big) - \frac{\partial}{\partial\sigma^{ab}}\big(S_0^{ab}\mu_t\big) \Big] \mathrm{d}\alpha\mathrm{d}\sigma  + \iint \hat{\tau}_{\alpha,\sigma} \Big[ \frac{1}{2} \frac{\partial^2}{\partial\alpha^a\partial\alpha^b}\big(S_{\mathrm{D}}^{ab} \mu_t\big) \Big] (\mathrm{d}\alpha\mathrm{d}\sigma) \label{eq:after-ibp}
\end{equation}
where we have used two steps of integration by parts in obtaining the last term.
No boundary terms will survive in \eqref{eq:after-ibp}: $\mu_t$ decays sufficiently rapidly as $|\alpha|\to\infty$, and its $\sigma$-support lies in the pure not-too-squeezed set $S_{\mathrm{NTS}}(\xi)$, whose eigenvalues are bounded above by $(\hbar/2)\xi^{-1}$; the normal flux through the NTS boundary vanishes by the NTS-preservation property established in Lemmas~6.1 and~7.1 of \cite{Hernandez2025}. Identifying the integrand with \(\hat{\mathcal{L}}^{(\alpha)}[\hat{\tau}_{\alpha,\sigma}]\mu_t\) and comparing the expressions, we arrive at the transport equation for \(\mu_t\):
\begin{equation}
\partial_t \mu_t(\alpha,\sigma) = -\frac{\partial}{\partial \alpha^a}\big( U^a \mu_t \big)
- \frac{\partial}{\partial \sigma^{ab}}\big( S_0^{ab} \mu_t \big)
+ \frac{1}{2} \frac{\partial^2}{\partial \alpha^a \partial \alpha^b}\big( S_{\mathrm{D}}^{ab} \mu_t \big).
\label{eq:mu-evolution}
\end{equation}
In contrast to \eqref{eq:naive-transport}, the contribution from \(S_{\mathrm{D}}\) appears as a diffusion term in \(\alpha\)-space rather than as a drift in \(\sigma\)-space.

At this point a comment on the meaning of \eqref{eq:mu-evolution} is in order. The measure defined by the push-forward \eqref{eq:pushforward} satisfies the transport equation \eqref{eq:naive-transport}. Equation \eqref{eq:mu-evolution} is the equation for an equivalent measure in which the part $S_D$ of the covariance drift has been converted into $\alpha$-diffusion with the help of \eqref{eq:sigma-alpha-relation}. This is the representation used in \cite{Hernandez2025} to keep the covariance matrix under control. For harmonic dynamics the two measures represent the same state, and below we work with the Fokker--Planck representation. For a single initial Gaussian the push-forward measure is a Dirac measure, while the equivalent Fokker--Planck measure spreads in $\alpha$; both give the same Gaussian state.

One may equivalently choose to shift a part of the \(\alpha\)-diffusion into the drift by writing
\begin{equation}
\frac{1}{2} \frac{\partial^2}{\partial\alpha^a\partial\alpha^b}(S_{\mathrm{D}}^{ab}\mu_t)
= \frac{\partial}{\partial\alpha^a}\Big( \frac{1}{2}\frac{\partial S_{\mathrm{D}}^{ab}}{\partial\alpha^b}\mu_t \Big) + \frac{\partial}{\partial\alpha^a}\Big( \frac{1}{2} S_{\mathrm{D}}^{ab} \frac{\partial\mu_t}{\partial\alpha^b} \Big),
\end{equation}
so that an effective drift \(\widetilde{U}^a = U^a + \frac{1}{2}\frac{\partial S_{\mathrm{D}}^{ab}}{\partial\alpha^b}\) can be defined (cf.~Eq.~(5.12) of \cite{Hernandez2025}). Note the plus sign: The spurious drift induced by the diffusion adds to the deterministic drift.

Equation \eqref{eq:mu-evolution} is a linear Fokker--Planck equation on the extended phase space \((\alpha,\sigma)\). The first two terms are drift terms representing the deterministic flow of the centroid and the non-diffusive part of the covariance dynamics. The last term is a diffusion term in \(\alpha\)-space, with diffusion matrix \(S_{\mathrm{D}}^{ab}\), which originates from the portion of the covariance evolution that has been converted into \(\alpha\)-diffusion via the identity \eqref{eq:sigma-alpha-relation}. Because \(S_{\mathrm{D}}\) is positive semidefinite, the equation preserves the nonnegativity of \(\mu_t\), and total probability is conserved under the usual regularity conditions on the drift and diffusion coefficients.

\section{Connection to the Generalized Ehrenfest Theorem}
\label{subsec:generalized-ehrenfest}

With the Gaussian mixture representation and the evolution equations for the individual components and for \(\mu_t\) in place, we can compute expectation values of arbitrary observables.

For any observable \(\hat{O}\) with Weyl symbol \(O(\beta)\), its expectation value in the approximate state \(\tilde{\rho}(t)\) is
\begin{equation}
\label{eq:expectation-general}
\langle \hat{O} \rangle_{\tilde{\rho}(t)} = \operatorname{Tr}\big( \tilde{\rho}(t) \hat{O} \big)
= \iint_{\mathbb{R}^{2d}\times\mathcal{S}} \mu_t(\alpha,\sigma) \; \langle \hat{O} \rangle_{\tau_{\alpha,\sigma}} (\mathrm{d}\alpha\,\mathrm{d}\sigma),
\end{equation}
where $\langle \hat{O} \rangle_{\tau_{\alpha,\sigma}}$ is defined by \eqref{999}.
Equation \eqref{eq:expectation-general} has a clear physical interpretation: the expectation value is obtained by averaging \(\langle\hat{O}\rangle_{\tau_{\alpha,\sigma}}\) over all Gaussian components with the probability measure \(\mu_t\). This turns the quantum dynamics into a classical statistical mechanics problem on the extended phase space \((\alpha,\sigma)\).

The generalized Ehrenfest theorem derived by Vallejo \textit{et al.} \cite{Vallejo2025} provides a universal decomposition of the time derivative of the expectation value of any observable \(\hat{O}\) in an open quantum system:
\begin{equation}
\label{eq:vallejo-general}
\frac{d\langle \hat{O}\rangle}{dt} = \sum_j \frac{d\lambda_j}{dt} \langle \psi_j | \hat{O} | \psi_j \rangle
+ \Big\langle \frac{\partial \hat{O}}{\partial t} \Big\rangle + \frac{i}{\hbar} \langle [\hat{\Omega}, \hat{O}] \rangle,
\end{equation}
where \(\rho(t) = \sum_j \lambda_j |\psi_j\rangle\langle\psi_j|\) is the spectral decomposition of a density matrix, and \(\hat{\Omega}\) is the Hermitian generator of the eigenbasis dynamics. (In \cite{Vallejo2025} the units $\hbar=1$ are used, so the last term there reads $i\langle[\hat{\Omega},\hat{O}]\rangle$.) The first term accounts for changes in the populations \(\lambda_j\) associated with entropy production, the second for explicit time dependence of \(\hat{O}\), and the third for coherent evolution. (A similar separation of time-dependence can be found in the generalization of Milburn's intrinsic decoherence model \cite{Milburn1991} to time-dependent Hamiltonians by Rajagopal \cite{Rajagopal1996}, who showed that an intrinsic time-dependence of the density matrix leads to an additive contribution in the rate of change of any expectation value.)

We now show how this decomposition emerges naturally from the Gaussian mixture framework.
Since $\langle \hat{O}\rangle_{\tau_{\alpha,\sigma}}$ has no explicit time dependence at fixed $(\alpha,\sigma)$, differentiating \eqref{eq:expectation-general} under the integral sign gives
\begin{equation}
\frac{d}{dt} \langle \hat{O} \rangle_{\tilde{\rho}(t)}
= \iint \big( \partial_t \mu_t \big) \langle \hat{O} \rangle_{\tau_{\alpha,\sigma}} \, \mathrm{d}\alpha\mathrm{d}\sigma. \label{eq:dt-expectation}
\end{equation}
The derivative along the semiclassical path is computed using \eqref{eq:alpha-evol} and \eqref{eq:sigma-evol}:
\begin{align}
\frac{d}{dt} \langle \hat{O} \rangle_{\tau_{\alpha,\sigma}}
&= U^a \partial_{\alpha^a} \langle \hat{O} \rangle_{\tau_{\alpha,\sigma}}
+ S^{ab} \partial_{\sigma^{ab}} \langle \hat{O} \rangle_{\tau_{\alpha,\sigma}}. \label{eq:dt-single}
\end{align}
Using the identity \(\partial_{\sigma^{ab}} \langle \hat{O} \rangle_{\tau_{\alpha,\sigma}} = \frac12 \partial_{\alpha^a}\partial_{\alpha^b} \langle \hat{O} \rangle_{\tau_{\alpha,\sigma}}\) (which follows from \eqref{eq:sigma-alpha-relation}) and the decomposition \(S=S_0+S_D\), we rewrite this as
\begin{align}
\frac{d}{dt} \langle \hat{O} \rangle_{\tau_{\alpha,\sigma}}
&= U^a \partial_{\alpha^a} \langle \hat{O} \rangle_{\tau_{\alpha,\sigma}}
+ S_0^{ab} \partial_{\sigma^{ab}} \langle \hat{O} \rangle_{\tau_{\alpha,\sigma}}
+ \frac12 S_D^{ab} \partial_{\alpha^a}\partial_{\alpha^b} \langle \hat{O} \rangle_{\tau_{\alpha,\sigma}}. \label{eq:dt-single-decomp}
\end{align}
By the weak transport equation \eqref{eq:weak-transport}, the right-hand side of \eqref{eq:dt-single-decomp} is the same total derivative as \eqref{eq:dt-expectation}; it is not an additional contribution, and the two must not be added.

Now substitute the Fokker--Planck equation \eqref{eq:mu-evolution} into \eqref{eq:dt-expectation}:
\begin{equation}
\iint \big( \partial_t \mu_t \big) \langle \hat{O} \rangle \mathrm{d}\alpha\mathrm{d}\sigma
= \iint \Big[ -\frac{\partial}{\partial\alpha^a}(U^a \mu_t) - \frac{\partial}{\partial\sigma^{ab}}(S_0^{ab} \mu_t) + \frac12 \frac{\partial^2}{\partial\alpha^a\partial\alpha^b}(S_D^{ab} \mu_t) \Big] \langle \hat{O} \rangle (\mathrm{d}\alpha\mathrm{d}\sigma).
\end{equation}
Integrating by parts for each term and assuming vanishing boundary contributions, we have, respectively
\begin{align}
-\iint \frac{\partial}{\partial\alpha^a}(U^a \mu_t) \langle \hat{O} \rangle &= \iint \mu_t\, U^a \partial_{\alpha^a} \langle \hat{O} \rangle, \label{eq:ibp1}\\
-\iint \frac{\partial}{\partial\sigma^{ab}}(S_0^{ab} \mu_t) \langle \hat{O} \rangle &= \iint \mu_t\, S_0^{ab} \partial_{\sigma^{ab}} \langle \hat{O} \rangle, \label{eq:ibp2}\\
\iint \frac12 \frac{\partial^2}{\partial\alpha^a\partial\alpha^b}(S_D^{ab} \mu_t) \langle \hat{O} \rangle &= \iint \mu_t\, \frac12 S_D^{ab} \partial_{\alpha^a}\partial_{\alpha^b} \langle \hat{O} \rangle. \label{eq:ibp3}
\end{align}
Collecting the terms, the total time derivative becomes
\begin{equation}
\begin{aligned}
\frac{d}{dt} \langle \hat{O} \rangle_{\tilde{\rho}(t)}
&= \iint \mu_t \Big( U^a \partial_{\alpha^a} \langle \hat{O} \rangle_{\tau_{\alpha,\sigma}}
+ S_0^{ab} \partial_{\sigma^{ab}} \langle \hat{O} \rangle_{\tau_{\alpha,\sigma}} \Big) (\mathrm{d}\alpha\mathrm{d}\sigma) \\
&\quad + \frac12 \iint \mu_t\, S_D^{ab} \partial_{\alpha^a}\partial_{\alpha^b} \langle \hat{O} \rangle_{\tau_{\alpha,\sigma}} (\mathrm{d}\alpha\mathrm{d}\sigma).
\end{aligned}
\label{eq:final-decomp}
\end{equation}

We now interpret the two integrals. The combination \(\big(U^a \partial_{\alpha^a} + S_0^{ab} \partial_{\sigma^{ab}}\big) \langle \hat{O} \rangle_{\tau_{\alpha,\sigma}}\) is the derivative generated by the symplectic (purity-preserving) part of the local harmonic evolution. In the frictionless case ($G=0$, $\Gamma=0$), this flow is generated by the quadratic Hamiltonian of the local harmonic approximation, so there exists a component-dependent Hermitian operator \(\hat{\Omega}_{\alpha,\sigma}\) (with units of energy in the convention of \eqref{eq:dt-tau-decomposed}) such that
\begin{equation}
U^a \partial_{\alpha^a} \hat{\tau}_{\alpha,\sigma} + S_0^{ab} \partial_{\sigma^{ab}} \hat{\tau}_{\alpha,\sigma} = -\frac{i}{\hbar} [\hat{\Omega}_{\alpha,\sigma}, \hat{\tau}_{\alpha,\sigma}].
\end{equation}
Consequently, the first integral equals the mixture average of $\frac{i}{\hbar}\langle [\hat{\Omega}_{\alpha,\sigma}, \hat{O}] \rangle$, which is the phase-space counterpart of the coherent term in \eqref{eq:vallejo-general}. It reduces to a single $\frac{i}{\hbar}\langle[\hat{\Omega},\hat{O}]\rangle$ when the generator is component-independent. When $G\neq0$ the drift contains a friction force that is not a Hamiltonian gradient in general, and no such $\hat{\Omega}_{\alpha,\sigma}$ exists; the first integral then also contains a dissipative contribution.

The second integral in \eqref{eq:final-decomp} involves the second derivatives of \(\langle \hat{O} \rangle\) and the diffusion matrix \(S_D\). It is the phase-space counterpart of the population-change term
\[
\sum_j \frac{d\lambda_j}{dt}\, \langle \psi_j|\hat{O}|\psi_j\rangle :
\]
it is controlled by the non-unitary part of the dynamics and vanishes for observables whose expectation value is linear in $\alpha$. A term-by-term identification with the eigenvalue derivative would require diagonalizing \(\tilde{\rho}(t)\) and is not needed here. This structure parallels the separation obtained by Rajagopal \cite{Rajagopal1996} for time-dependent Hamiltonians, where the intrinsic time-dependence of the density matrix gives rise to an additive contribution to $d\langle\hat{A}\rangle/dt$ that cannot be absorbed into the Heisenberg commutator. In our framework, that same physics is encoded in the \(S_D\) diffusion term, thereby providing a phase-space origin for the entropy-producing part of the generalized Ehrenfest theorem.

%Finally, any explicit time dependence of the observable \(\hat{O}\) contributes \(\langle \partial_t \hat{O} \rangle\), which may be added separately.

Thus, the Gaussian mixture framework not only reproduces the structure of the generalized Ehrenfest theorem but also provides a transparent microscopic picture that  coherent motion is encoded in the centroid and symplectic covariance trajectories, while decoherence and dissipation manifest themselves through the diffusive reshaping of the mixing measure.

For the special choice $\hat{O} = \hat{H}$ the decomposition \eqref{eq:final-decomp} acquires a transparent thermodynamic meaning.
The first integral,
\begin{equation}\label{eq:work-def}
\dot{W} := \iint \mu_t \Big( U^a \partial_{\alpha^a} \langle \hat{H} \rangle_{\tau_{\alpha,\sigma}}
+ S_0^{ab} \partial_{\sigma^{ab}} \langle \hat{H} \rangle_{\tau_{\alpha,\sigma}} \Big) \mathrm{d}\alpha\mathrm{d}\sigma,
\end{equation}
involves the drift $U^a$ and the symplectic covariance flow $S_0^{ab}$; in the frictionless case the drift is the Hamiltonian drift $\partial^a H$. It describes how the macroscopic centroid motion and the symplectic squeezing of each wave packet change the energy without altering the statistical weights. In the frictionless harmonic case it represents the reversible part of the energy change and vanishes for a time-independent Hamiltonian, since the unitary flow conserves the energy. The second integral,
\begin{equation}\label{eq:heat-def}
\dot{Q} := \frac12 \iint \mu_t\, S_D^{ab} \partial_{\alpha^a}\partial_{\alpha^b} \langle \hat{H} \rangle_{\tau_{\alpha,\sigma}} \mathrm{d}\alpha\mathrm{d}\sigma,
\end{equation}
is driven solely by the diffusion matrix $S_D$. It captures the irreversible broadening of the phase-space distribution and, for a time-independent Hamiltonian in the frictionless harmonic case, is the entire energy change; it is then naturally identified with the heat absorbed by the system. With these definitions,
\begin{equation}
\dot{W} + \dot{Q} = \frac{d}{dt}\langle \hat{H}\rangle_{\tilde{\rho}(t)},
\label{eq:first-law}
\end{equation}
so that the first law is satisfied by construction.

\section{Example: Free Particle with Momentum Diffusion}
\label{sec:example}

We illustrate the above results by a simple analytically tractable example: a free particle in one dimension subject to a Lindblad operator that induces momentum diffusion (cf.~\cite{Graefe2018}).

Consider a one-dimensional system with Hamiltonian \(\hat{H} = \hat{p}^2/(2m)\) and a single Hermitian linear Lindblad generator \(\hat{L} = \sqrt{2\lambda}\,\hat{x}\), where \(\lambda > 0\) parameterizes the coupling strength to the environment. The Lindblad master equation reads
\begin{equation}
\partial_t \hat{\rho} = -\frac{i}{\hbar}[\hat{H},\hat{\rho}] + \frac{2\lambda}{\hbar}\big( \hat{x}\hat{\rho}\hat{x} - \frac{1}{2}\{\hat{x}^2,\hat{\rho}\} \big).
\label{eq:example-lindblad}
\end{equation}
This is the standard position-dephasing model: it generates pure momentum diffusion without friction. The mean position and momentum evolve classically, the momentum variance grows linearly in time, and the position second moment acquires a cubic-in-time contribution.

The Weyl symbols for the Hamiltonian $\hat{H}$ and the Lindblad generator are respectively
\begin{equation}
H(x,p) = \frac{p^2}{2m},\qquad L(x,p)=\sqrt{2\lambda}\,x.
\end{equation}
From Eqs.~\eqref{eq:drift}--\eqref{eq:S} we compute
\begin{equation}
U^x = \partial^x H = \frac{p}{m},\quad U^p = \partial^p H = 0,\quad G^a =0,
\end{equation}
\begin{equation}
D^{pp} = \hbar\, (\partial^p L)(\partial^p L^*) = 2\hbar\lambda,\qquad D^{xx}=D^{xp}=0,
\end{equation}
and
\begin{equation}
F = \begin{pmatrix} 0 & 1/m \\ 0 & 0 \end{pmatrix},\qquad \Gamma = 0.
\end{equation}
Here we used $\partial^x=\partial_p$ and $\partial^p=-\partial_x$. The diffusion is in the momentum direction, as it must be for a Lindblad operator that is diagonal in the position basis.
Thus the centroid obeys
\begin{equation}
\frac{dx}{dt} = \frac{p}{m},\qquad \frac{dp}{dt}=0
\label{eq:example-centroid}
\end{equation}
whose solution is
\begin{equation}
 x(t)=x_0 + \frac{p_0}{m}t,\quad p(t)=p_0
\label{eq:example-centroid-sol}
\end{equation}
with $(x_0,p_0)$ being the initial condition.
The covariance matrix evolves according to
\begin{equation}
\frac{d}{dt} \begin{pmatrix} \sigma^{xx} & \sigma^{xp} \\ \sigma^{xp} & \sigma^{pp} \end{pmatrix}
= \begin{pmatrix} \frac{2}{m}\sigma^{xp} & \frac{1}{m}\sigma^{pp} \\[2pt] \frac{1}{m}\sigma^{pp} & 2\hbar\lambda \end{pmatrix}. \label{eq:cov-free}
\end{equation}
For an initial minimum-uncertainty state \((x_0,p_0)\) with \(\sigma_0 = \frac{\hbar}{2} I\), the solution is
\begin{align}
\sigma^{pp}(t) &= \frac{\hbar}{2}+2\hbar\lambda t,\qquad
\sigma^{xp}(t) = \frac{\hbar}{2m}t+\frac{\hbar\lambda}{m}t^2,\nonumber\\
\sigma^{xx}(t) &= \frac{\hbar}{2} + \frac{\hbar}{2m^2}t^2 + \frac{2\hbar\lambda}{3m^2}t^3. \label{eq:cov-sol}
\end{align}

Because the Lindblad operator is linear and the Hamiltonian is quadratic, the local harmonic approximation is exact, and the decomposition \(S=S_0+S_D\) may be taken as \(S_D = D\) and \(S_0 = S-D\) (the frictionless case). The resulting Fokker--Planck equation for the mixing measure \(\mu_t\) (Eq.~\eqref{eq:mu-evolution}) reduces to a purely deterministic flow in the \(\sigma\) directions (with no \(\sigma\)-diffusion) plus an \(\alpha\)-diffusion in the momentum direction. For a single initial Gaussian, the push-forward measure is a Dirac measure on the solution curve \eqref{eq:cov-sol}; the equivalent Fokker--Planck representation spreads it in $\alpha$, but both representations yield the same Gaussian state. For mixtures of coherent states with fixed covariance, a rigorous long-time Fokker--Planck correspondence was proved in \cite{Galkowski2025}. Here, we consider general Gaussian states with evolving covariance and focus on the Ehrenfest decomposition. The covariance \eqref{eq:cov-sol} remains positive definite and satisfies the uncertainty principle, and its minimum eigenvalue stays above $0.25\,\hbar$ over the time interval shown, so the not-too-squeezed condition is maintained for a suitable $\xi\lesssim0.5$.

Let us focus on the observables \(\hat{O}=\hat{x}^2\) and \(\hat{O}=\hat{p}^2\). For a single Gaussian component,
\begin{equation}
\langle \hat{x}^2 \rangle_{\tau_{\alpha,\sigma}} = x(t)^2 + \sigma^{xx}(t),\qquad
\langle \hat{p}^2 \rangle_{\tau_{\alpha,\sigma}} = p(t)^2 + \sigma^{pp}(t).
\end{equation}
For an initial pure coherent state, the push-forward measure reduces to a Dirac measure, so the mixture expectation values reduce to
\begin{equation}
\langle \hat{x}^2 \rangle_{\tilde{\rho}(t)} = \left(x_0 + \frac{p_0}{m}t\right)^2 + \frac{\hbar}{2} + \frac{\hbar}{2m^2}t^2 + \frac{2\hbar\lambda}{3m^2}t^3,
\label{eq:x2-example}
\end{equation}
\begin{equation}
\langle \hat{p}^2 \rangle_{\tilde{\rho}(t)} = p_0^2 + \frac{\hbar}{2} + 2\hbar\lambda t.
\label{eq:p2-example}
\end{equation}
The time derivatives decompose as
\begin{align}
\frac{d}{dt} \langle \hat{x}^2 \rangle
&= \underbrace{\frac{2}{m}x p + \frac{2}{m}\sigma^{xp}}_{\text{coherent}} + \underbrace{0}_{\text{diffusive}}, \label{eq:dx2}\\
\frac{d}{dt} \langle \hat{p}^2 \rangle
&= \underbrace{0}_{\text{coherent}} + \underbrace{2\hbar\lambda}_{\text{diffusive}}. \label{eq:dp2}
\end{align}
For $\hat{x}^2$ the direct diffusion term vanishes because $\partial_p^2\langle \hat{x}^2\rangle=0$; the growth of $\langle \hat{x}^2\rangle$ is entirely due to the coherent covariance flow, which has already incorporated the diffusion through $\sigma^{xp}$. For $\hat{p}^2$ the coherent term vanishes and the diffusive term is directly $D^{pp}=2\hbar\lambda$, which makes $\hat{p}^2$ the cleanest illustration of the separation.

The energy is
\begin{equation}
\langle \hat{H} \rangle_{\tilde{\rho}(t)} = \frac{p_0^2}{2m} + \frac{\hbar}{4m} + \frac{\hbar\lambda}{m}t,
\end{equation}
so
\begin{equation}
\dot{Q} = \frac12 D^{pp}\partial_p^2 \langle \hat{H}\rangle = \frac{\hbar\lambda}{m} = \frac{d}{dt}\langle \hat{H}\rangle,
\end{equation}
and $\dot{W}=0$: the energy increase is purely heat, consistent with the time-independent Hamiltonian and the frictionless dynamics.

\subsection{Numerical verification}
\label{subsec:numerics}
\begin{figure}[t]
\centering
\includegraphics[width=0.95\textwidth]{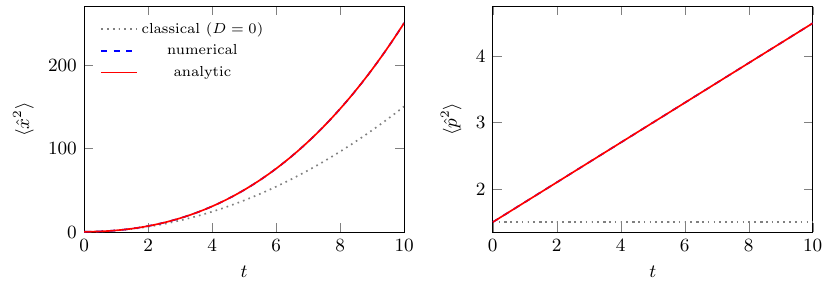}
\caption{Left: the second moment $\langle\hat{x}^2\rangle(t)=x(t)^2+\sigma^{xx}(t)$; right: the second moment $\langle\hat{p}^2\rangle(t)=p(t)^2+\sigma^{pp}(t)$. The blue dashed lines are the numerical Euler solution, the red solid lines the analytical solution, and the grey dotted lines the classical comparison with $D=0$. The quantum case uses $D^{ab}=\mathrm{diag}(0,2\hbar\lambda)$. The parameters are $m=1$, $\hbar=1$, $\lambda=0.15$, $x_0=0$, $p_0=1$, $\sigma_0=\frac{\hbar}{2}I_2$, $\delta t=0.002$, and $T_{\max}=10$.}
\label{fig:example}
\end{figure}

We integrate the example ODEs \eqref{eq:example-centroid} and \eqref{eq:cov-free} with a simple Euler scheme, $\delta t=0.002$ up to $T_{\max}=10$, and compare with the analytic solution \eqref{eq:example-centroid-sol} and \eqref{eq:cov-sol}. The parameters are $m=1$, $\hbar=1$, $\lambda=0.15$, $x_0=0$, $p_0=1$, and $\sigma_0=\frac{\hbar}{2}I_2$. The relative error of $\langle\hat x^2\rangle$ is $7.8\times10^{-4}$ and that of $\langle\hat p^2\rangle$ is below $10^{-13}$. Figure~\ref{fig:example} shows the resulting second moments and the classical $D=0$ comparison. The quantum curve acquires the cubic term $2\hbar\lambda t^3/(3m^2)$ and the linear term $2\hbar\lambda t$ derived above.

\begin{figure}[h!]
\centering
\includegraphics[width=0.95\textwidth]{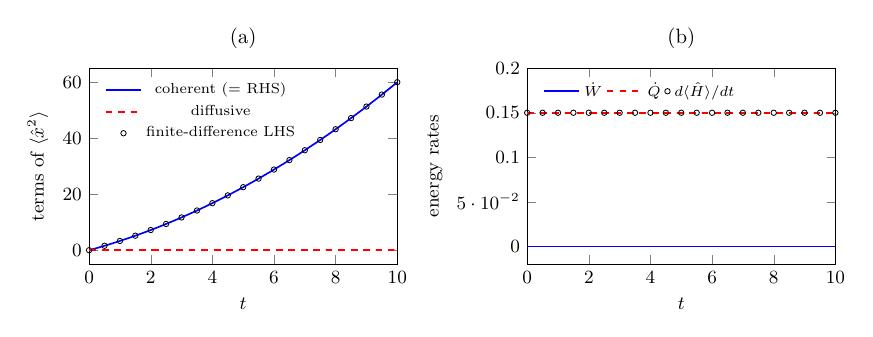}
\caption{(a) Ehrenfest decomposition \eqref{eq:final-decomp} for $\hat O=\hat x^2$ in the free-particle example: coherent term (blue solid; this is the right-hand side, since the diffusive term vanishes), diffusive term (red dashed), and the finite-difference derivative of \eqref{eq:x2-example} (black circles). (b) First law \eqref{eq:first-law} for $\hat O=\hat H$: $\dot W$ (blue solid), $\dot Q$ (red dashed), and $d\langle\hat H\rangle/dt$ (black circles). }
\label{fig:example-ehrenfest}
\end{figure}

For the same example, the decomposition \eqref{eq:final-decomp} reduces to \eqref{eq:dx2}--\eqref{eq:dp2}. We evaluate these terms from the analytic solution and compare with finite-difference derivatives of \eqref{eq:x2-example}--\eqref{eq:p2-example}. For $\hat x^2$ the maximum residual is $8.0\times10^{-7}$ and for $\hat p^2$ it is $6.4\times10^{-13}$. For $\hat O=\hat H$, \eqref{eq:work-def} and \eqref{eq:heat-def} give $\dot W=0$ and $\dot Q=\hbar\lambda/m$. Figure~\ref{fig:example-ehrenfest}(b) compares $\dot W$, $\dot Q$, and $d\langle\hat H\rangle/dt$. The first law \eqref{eq:first-law} is satisfied with a maximum residual of $3.2\times10^{-13}$. Figure~\ref{fig:example-ehrenfest}(a) shows the two terms for $\hat x^2$, their sum, and the finite-difference derivative of \eqref{eq:x2-example}.

\FloatBarrier
\section{Conclusion}
\label{sec:conclusion}
We have presented a semiclassical kinetic description of open quantum systems that rests on the rigorous mixture approach of \cite{Hernandez2025}.  By making the Fokker--Planck equation for the mixing measure \(\mu_t\) explicit and by connecting it to the generalized Ehrenfest theorem, we provide a transparent phase-space decomposition of expectation-value dynamics into coherent and incoherent parts.   A simple analytically solvable example illustrates the general results, from the semiclassical path equations to the Ehrenfest decomposition; the example also shows explicitly that the diffusion term can affect an observable indirectly through the covariance flow, so that the split between the two contributions depends on the observable. The numerical checks confirm the example's analytic solution, the Ehrenfest decomposition, and the first law.

\section*{Acknowledgements}
 The author would like to acknowledge the support from YCIT(xjr2024030).

\end{document}